# The Parent of Misfit-Layered Cobalt Oxides: $[Sr_2O_2]_qCoO_2$

H. Yamauchi[1], K. Sakai[1], T. Nagai[2], Y. Matsui[2], and M. Karppinen[1,*]

[1]*Materials and Structures Laboratory, Tokyo Institute of Technology, 4259 Nagatsuta, Midori-ku, Yokohama 226-8503, Japan*

[2]*National Institute for Materials Science, 1-1 Namiki, Tsukuba, Ibaraki 305-0044, Japan*

Misfit-layered (ML) cobalt oxides of the general formula of $[M_mA_2O_{m+2}]_qCoO_2$ have been proven to be efficient thermoelectric materials as the structure is capable in accommodating the two seemingly contradictory characteristics of high electrical conductivity and large thermo-electric power. They are also potential hosts for other *oxymoron*-like functions. The known phases all contain one or two square-planar $M$O ($M$ = Co, Bi, Pb, Tl, *etc*.) layers sandwiched together with $A$O ($A$ = Ca, Sr, Ba, *etc*.) planes of square symmetry and $CoO_2$ layers of hexagonal symmetry. Here we report realization of the simplest ($m$ = 0) ML phase forming in the Sr-Co-O system with the cation ratio, Sr/Co = 1. Atomic-resolution TEM imaging confirms for the new phase the parent three-layer crystal structure, SrO-SrO-$CoO_2$, which is compatible with the formula of $[Sr_2O_2]_qCoO_2$. Electron diffraction reveals that the phase is rather commensurate, *i.e.* the "misfit parameter" $q$ is 0.5. Nevertheless, in terms of the transport-property characteristics the new ML parent is comparable to its earlier-established and more complex derivatives.

Keywords:

Misfit-layered oxide, Cobalt oxide, Thermoelectric oxide, New-material design, Electron microscopy

**Introduction**

"Layer-engineered" oxides of the 3$d$ transition metals have been widely recognized as candidates for the next-generation electronics materials. The phases derived from the $CuO_2$ layer and exhibiting high-$T_c$ superconductivity form one of the most impressive families of such oxides, in terms of both the variety of members and the technological impact. Another family of multi-layered oxides of high promises was discovered more recently, that is, the "misfit-layered" (ML) cobalt oxides [1]. In conventional multi-layered oxides such as the superconductive Cu oxides the individual layers are stacked to form a crystallographically coherent crystal, whereas in the ML compounds a hexagonal ($CdI_2$-structured) $CoO_2$ layer with triangular arrangement of the constituent atoms, O or Co, in each sub-layer is coupled incoherently with a square-planar (rock-salt-type) $[(MO)_m(AO)_2]$ layer-block ($M$ = Co, Bi, Pb, Tl, *etc*.; $A$ = Ca, Sr, Ba, *etc*.). A schematic crystal structure is shown in Fig. 1. Each ML oxide repeats the layer sequence of $AO$-$(MO)_m$-$AO$-$CoO_2$ and obeys the formula of $[M_mA_2O_{m+2}]_qCoO_2$ ($q$: "misfit parameter", the value of which ranges within 0.50 ~ 0.62 for known ML oxides). The different blocks in the ML structure possess not only different crystal symmetries but are different in the chemical nature and electronic structure. This provides us with possibilities for incorporating multiple functions into a single material. The first such "combinatorial" function discovered for ML cobalt oxides is the unexpectedly good thermoelectric (TE) performance that originates from the fact that these compounds can play the dual role of being concomitantly both a poor thermal conductor and a good electrical conductor [2,3].

For the materials science community, an ultimate goal should always be to find the limits of each of newly-discovered material families. This has been the motivation for the new-material search within the superconductive copper-oxide family, $M_mA_2Q_{n-1}Cu_nO_{m+2+2n}$ ($M$ = Cu, Bi, Pb, Tl, Hg, *etc*.; $A$ = Ba, Sr, La, *etc*.; $Q$ = Ca, *etc*.) [4]: the family has been extended, in terms of $m$ up to 3 and of $n$ up to 9 [5,6], and also derived down to the parent $m$ = 0 or "zero" phases [7-11]. In a parallel manner, for the ML cobalt-oxide family too, the target should be to find the limits of material variety, from the simplest to the much more complex. Here we report discovery of the first zero ($m$ = 0) ML cobalt oxide expressed as $[Sr_2O_2]_qCoO_2$.

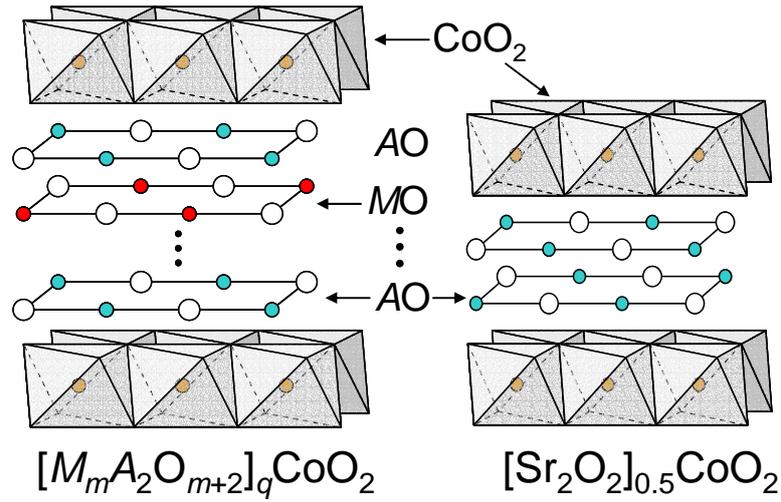

**Fig. 1.** Crystal structures of misfit-layered cobalt oxides, $[M_mA_2O_{m+2}]_qCoO_2$, in general and of the new "zero" phase, $[Sr_2O_2]_{0.5}CoO_2$. The former contains hexagonal $CoO_2$ layers coupled incoherently with square-planar $AO$ and $MO$ layers along the sequence, $AO$-$(MO)_m$-$AO$-$CoO_2$, whereas the latter lacks the $(MO)_m$ "charge reservoir". All the layers should be considered potentially non-stoichiometric at least in terms of oxygen.

## Experimental Section

We synthesized the Sr-Co-O samples in evacuated quartz ampoules at 850 °C from a mixture of $SrO_2$ (freshly prepared *prior* to use [12]) and $Co_3O_4$ powders with the ratio of 1:1 for the constituent metals, Sr and Co. In terms of oxygen, the precursor mixture serves as a source for moderately oxidizing conditions as it corresponds to the oxygen-excess nominal composition of "$[Sr_2O_2]_{0.5}CoO_{2.33}$". Other cation ratios and synthesis temperatures were tested as well but found to yield less phase-pure samples. The phase composition was determined from x-ray powder diffraction (XRD) patterns collected at room temperature (Rigaku: RINT-2500V equipped with a rotating anode; Cu $K_\alpha$ radiation). The zero structure was confirmed from high-resolution transmission-electron microscopy (HRTEM) images and electron diffraction (ED) patterns (Hitachi: H-1500; acceleration voltage 820 kV). The actual chemical composition of the target phase was determined by energy dispersive x-ray spectroscopy (EDS) analyzer attached to a TEM microscope (Hitachi: HF-3000S; acceleration voltage 300 kV).

Lattice parameter refinement was carried out on the basis of the XRD data using the Rietveld refinement program JANA2000. Electrical resistivity was measured for the samples in the temperature range of 4 ~ 350 K by a 4-probe technique (Quantum Design: PPMS), and magnetic properties were evaluated from the data collected from 2 to 300 K with a superconducting-quantum-interference-device (SQUID) magnetometer (Quantum Design: MPMS-XL; field-cooled mode) under 100 Oe. The thermoelectric power was measured in the temperature range of 5 ~ 280 K with a steady-state technique.

## Results and Discussion

Our ampoule synthesis yielded polycrystalline samples that were not completely free from the starting materials: small peaks due to $Co_3O_4$ and $SrCO_3$ (rather than $SrO_2$) were always seen in the XRD patterns recorded for the samples (Fig. 2). Apparently during the course of the synthesis procedure the non-yet-reacted $SrO_2$ readily transforms into $SrCO_3$. Here we should mention that this happened despite our best efforts to avoid carbon contamination; we for instance carefully checked that the $SrO_2$ powder used for the synthesis was free from $SrCO_3$ within the XRD detection limit. After subtracting the contributions from the apparent $Co_3O_4$ and $SrCO_3$ impurity phases, the remaining diffraction lines in the XRD patterns could not be explained for any known compound(s). Surprisingly, no signs of the two most common ternary phases in the Sr-Co-O system, *i.e.* the brownmillerite $Sr_2Co_2O_5$ and the perovskite $SrCoO_{3-\delta}$, were seen. The former phase is the one obtained by synthesis carried out in air [13], whereas the latter forms through high-pressure (~6 GPa, $KClO_4$ as an oxygen source) [14] or electrochemical [15] oxidation. Apparently we have successfully attained a not-yet-combed intermediate oxygen-pressure range which does not allow stabilization of these already known Sr/Co = 1 phases.

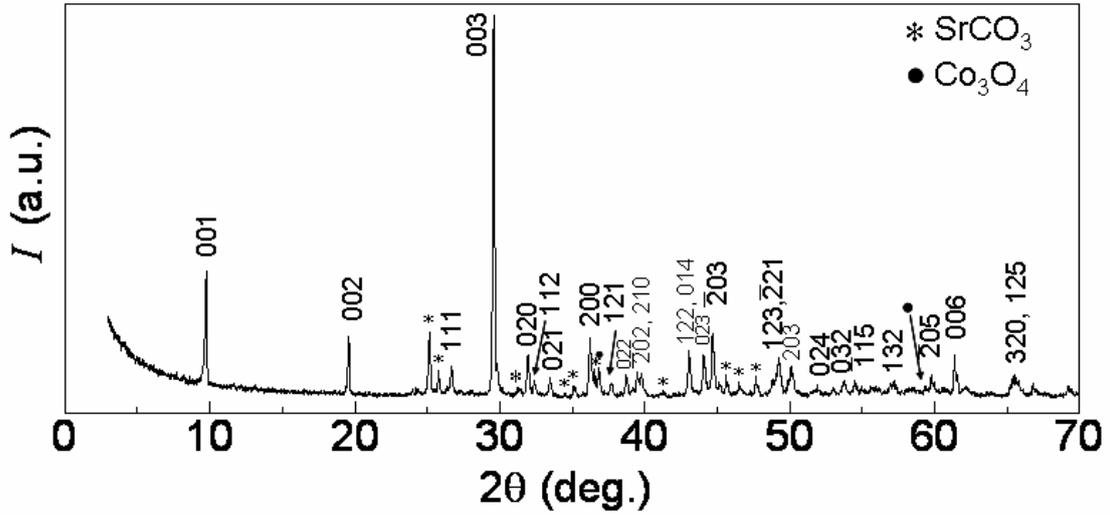

**Fig. 2.** X-ray powder diffraction pattern for a sample of the new $[Sr_2O_2]_{0.5}CoO_2$ phase. Besides the reflections due to the precursors, $SrO_2$ and $Co_3O_4$, all the remaining peaks can be indexed according to the structure model expected for a "zero" ML cobalt oxide phase as sketched in Fig. 1 ($P2_1/m$; $a = 4.980$ Å, $b = 5.596$ Å, $c = 9.107$ Å, $\beta = 96.28°$).

We utilized HRTEM for the first identification of the new phase (Fig. 3a). The HRTEM image revealed a perfectly arranged (2+1)-layer structure for the cations (*i.e.* heavier atoms) compatible with that of the "zero phase", $[Sr_2O_2]_qCoO_2$. The layer-repetition thickness was evaluated on the basis of the HRTEM image at ~9 Å. The HRTEM image furthermore suggested monoclinic symmetry, similar to that seen for known ML cobalt oxides. The actual phase-specific chemical composition was determined by TEM-EDS at Sr/Co = 0.99(5) using signals from several different grains.

The monoclinic distortion revealed from the HRTEM image was verified from the ED pattern taken with the electron beam along [010] direction, *i.e.* angle $\angle ac \approx 96°$ (Fig. 3b), whereas angle $\angle bc$ was confirmed to be 90° (Fig. 3c). From the ED pattern taken with the electron beam along [001] direction (Fig. 3d) it was clearly revealed that the new phase is commensurate, *i.e.* the hexagonal (H) and square (S) lattices coincide such that the misfit parameter $q = b_H/b_S$ for $[Sr_2O_2]_qCoO_2$ is 0.5. In Fig. 3d, only ($0k0$) reflections with $k$ = even appear, whereas in Fig. 3c reflections with $k$ = odd are seen as well. In the latter case, the $k$ = odd reflections are most likely induced by multiple scattering effects [16]. Accordingly, on the basis of the ED data, the lattice structure of the new Sr-Co-O compound could be derived as follows: monoclinic space group $P2_1/m$ (No. 11) [16], lattice parameters $a \approx 5.0$ Å, $b \approx 5.7$ Å, $c \approx 9.1$ Å, $\beta \approx 96°$. The more precise lattice parameters were then refined from the XRD data at: $a = 4.980(0.1)$ Å, $b = 5.596(0.1)$ Å, $c = 9.107(0.1)$ Å, $\beta = 96.28(0.1)°$. These results are highly consistent with our imagination (Fig. 1) of the crystal structure of the new phase, that is, a commensurate zero ML cobalt oxide with the layer sequence of SrO-SrO-CoO$_2$.

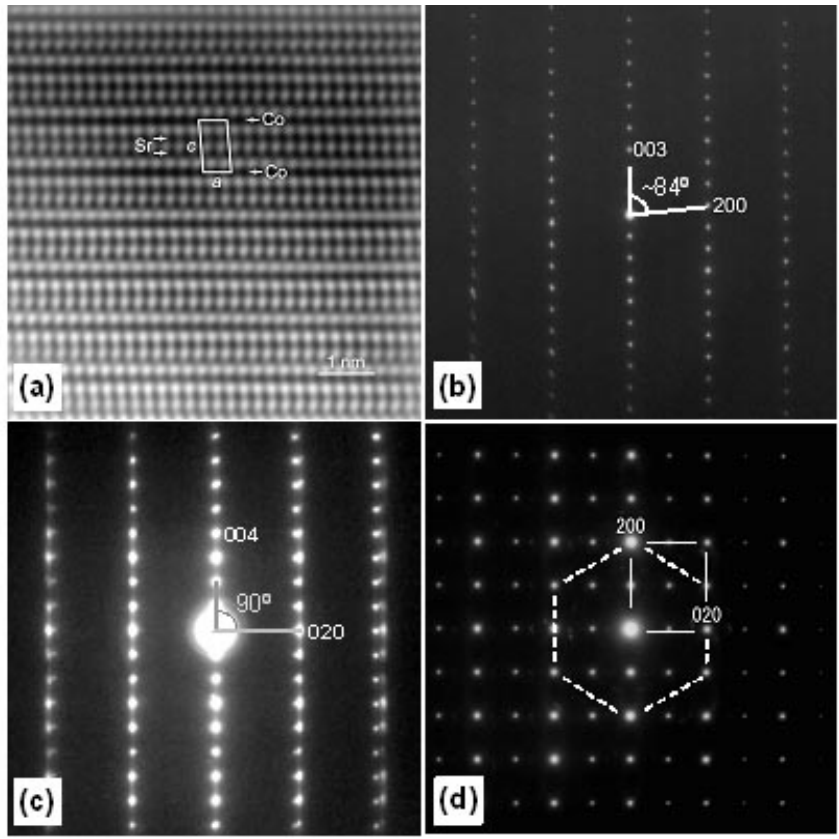

**Fig. 3.** HRTEM image and ED patterns for $[Sr_2O_2]_{0.5}CoO_2$. (a) The HRTEM image represents the *ac* plane, exhibiting the layer sequence of SrO-SrO-CoO$_2$ and a monoclinic distortion. The ED patterns are taken with the electron beam (b) along [010], (c) [100] and (d) [001]. The first ED pattern confirms the monoclinic distortion and the last shows that the phase is commensurate, i.e. $q\,(=b_H/b_S) = 0.5$.

As for the observed $q$ value of 0.5, a similar commensurate ML structure has for the first time been observed for $[Bi_2Ba_2O_4]_{0.5}CoO_2$ [17]. Apparently the large size of the *A*-site cation (= Ba) in $[Bi_2Ba_2O_4]_{0.5}CoO_2$ allows the structure to release the drive for misfitness. In the present case too, the *ab* plane is considerably expanded in comparison to the other known Sr-based ML cobalt oxides to rather become close to that for $[Bi_2Ba_2O_4]_{0.5}CoO_2$ (see Table 1), thus rationalizing the low (and commensurate) value of 0.5 for *q* in $[Sr_2O_2]_qCoO_2$. As a plausible explanation for the larger-than-expected *ab*-plane dimension we suggest that the $[Sr_2O_2]_{0.5}CoO_2$ phase accommodates considerable concentration of oxygen vacancies. From the redox chemistry point of view, this is highly reasonable since without oxygen vacancies the valence of cobalt in $[Sr_2O_2]_{0.5}CoO_2$ would be as high as +4 (*cf.* $SrCoO_3$ with the same nominal chemical composition). From previous experiences, such a high value of cobalt valence is achieved under highly oxidizing conditions only. The present synthesis conditions are very similar to those employed for other ML and related cobalt oxides, for which the valence of cobalt remains well below +3.5 [18]. Actually, even the non-zero ML cobalt oxides have been shown to be prone to oxygen vacancies [18]. We should also recall corresponding copper-oxide zero-phase, $Sr_2CuO_{4-\delta}$ which is believed to be strongly oxygen-deficient [19].

**Table 1.** Lattice-parameter data for $[Sr_2O_2]_{0.5}CoO_2$, some representative ("non-zero") ML cobalt oxides of $[M_mA_2O_{m+2}]_qCoO_2$, the first-generation oxide thermoelectrics $Na_{0.74}CoO_2$ and its superconductive water-derivative $Na_{0.35}CoO_2 \bullet 1.3H_2O$ as well as ion-exchanged Sr-counterpart $Sr_{0.35}CoO_2$. It is interesting to note that the $CoO_2$-layer separation distance in $[Sr_2O_2]_{0.5}CoO_2$ is close to that in $Na_{0.35}CoO_2 \bullet 1.3 H_2O$, but the new phase does not exhibit superconductivity (above 2 K).

| Phase | $q$ | $a$ [Å] | $b_S$ [Å] | $b_H$ [Å] | $c$ [Å] | β | Ref. |
|---|---|---|---|---|---|---|---|
| $[Sr_2O_2]_qCoO_2$ | 0.50 | 5.0 | 5.6 | 2.8 | 9.1 | 96.2 | present |
| $[Bi_2Ba_2O_4]_qCoO_2$ | 0.50 | 5.0 | 5.6 | 2.8 | 15.5 | 92.0 | 17 |
| $[Bi_2Sr_2O_4]_qCoO_2$ | 0.56 | 4.9 | 5.1 | 2.8 | 14.9 | 93.5 | 20 |
| $[Bi_2Ca_2O_4]_qCoO_2$ | 0.60 | n/a | n/a | n/a | n/a | n/a | 21 |
| $[CoCa_2O_3]_qCoO_2$ | 0.62 | 4.8 | 4.6 | 2.8 | 10.8 | 98.1 | 22 |
| $[(Co,Cu)_2Ca_2O_3]_qCoO_2$ | 0.62 | 4.8 | 4.5 | 2.8 | 12.8 | 93.9 | 23 |
| $Na_{0.74}CoO_2$ | - | 5.6 | - | 2.8 | 5.4 | - | 24 |
| $Na_{0.35}CoO_2 \bullet 1.3 H_2O$ | - | 5.6 | - | 2.8 | 9.8 | - | 25 |
| $Sr_{0.35}CoO_2$ | - | 5.6 | - | 2.8 | 5.8 | - | 26 |

We performed post-annealing experiments in order to check whether it would be possible to control the concentration of oxygen vacancies in $[Sr_2O_2]_{0.5}CoO_2$. Reductive annealing carried out in $N_2$ gas at 400 °C was found to decrease the *c* lattice parameter from ~9.11 Å of the as-synthesized sample to ~9.00 Å of the annealed sample. Moreover, since the $N_2$-annealing was performed in a thermobalance the amount of removable oxygen could be estimated to be 0.10 ~ 0.15 oxygen atoms *per* formula unit. It thus seems that $[Sr_2O_2]_{0.5}CoO_2$ possesses oxygen vacancies of a concentration that is tunable to some extend. Attempts to increase oxygen content were made as well by annealing specimens of the as-synthesized material in a cubic-anvil high-pressure apparatus at 5 GPa and 400 °C in the presence of $KClO_3$ as an excess-oxygen source. However, only a slight increase in the *c* parameter from ~9.11 Å to ~9.12 Å was observed.

The thermoelectric power data (Fig. 4a) for the as-synthesized material are consistent with our tentative suggestion that $[Sr_2O_2]_{0.5}CoO_2$ is oxygen-deficient: the Seebeck coefficient (*S*) is positive (at temperatures above ~40 K) indicating that the majority carriers are holes such that $[Sr_2O_2]_{0.5}CoO_2$ is a hole-doped $Co^{III}$ lattice rather than an electron-doped $Co^{IV}$ lattice. From Fig. 4a it should also be noted that the absolute value of *S* at room temperature is as high as ~85 μV/K, being comparative to those reported for various other ML and related oxides based on the $CdI_2$-structured $CoO_2$ layer [2,3]. The cause of the anomalous behaviour seen in the *S versus* temperature (*T*) curve at low temperatures is unknown, though it seems to coincide with the anomaly in the magnetic susceptibility (χ) *versus T* curve (Fig. 4b), which is most likely due to the antiferromagnetic ($T_N \approx 33$ K [27]) $Co_3O_4$ impurity present in the sample. The high-temperature (*T* > 35 K) portion of the curve looks rather levelled off but further analyses are not appropriate because of the superimposed effect from the impurity.

Figure 5 shows the *T* dependence of resistivity (ρ) for our best $[Sr_2O_2]_{0.5}CoO_2$ sample. The curve exhibits semiconducting behaviour within the whole temperature range measured (4 - 350 K). The absolute resistivity values are somewhat (one or two digits) higher than those typically reported for polycrystalline ML cobalt-oxide samples. This is likely due to the non-reacted precursor traces of $SrO_2$ and $Co_3O_4$ in the sample. Actually, at temperatures higher than ~17 K the ρ *versus T* data are well explained with the thermally-activated conduction mechanism: from the fitting to $\rho(T) = \rho_0 \exp(E_a/k_B T)$ (see the inset) the activation energy, $E_a$, was found as low as ~7 meV.

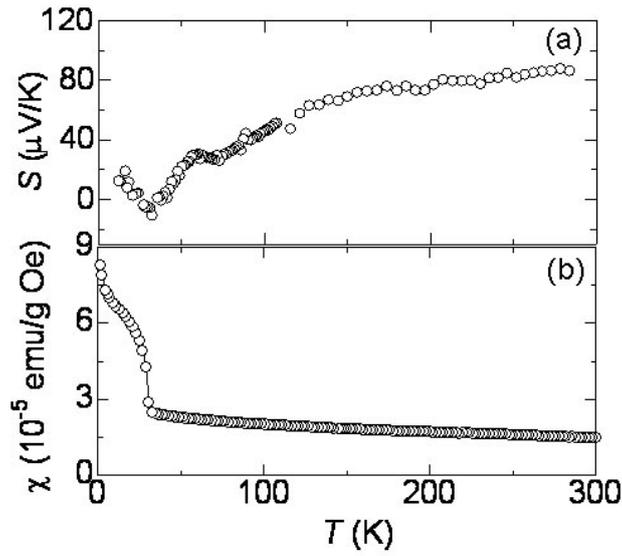

**Fig. 4.** Temperature ($T$) dependence of (a) Seebeck coefficient ($S$), and (b) magnetic susceptibility ($\chi$) of $[Sr_2O_2]_{0.5}CoO_2$.

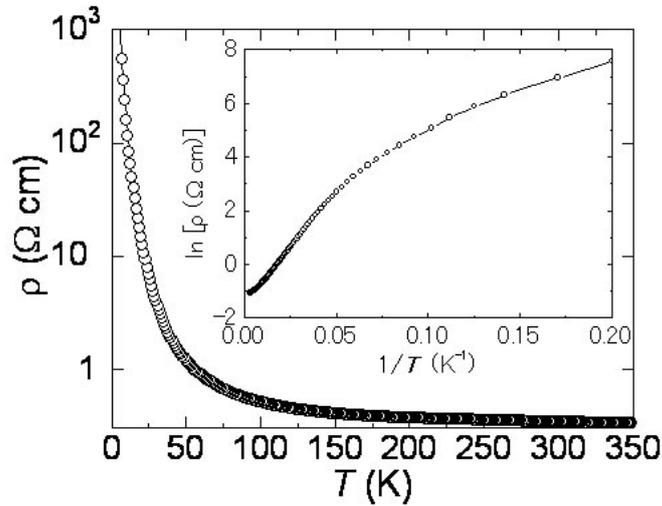

**Fig. 5.** Resistivity ($\rho$) *versus* temperature ($T$) characteristics of $[Sr_2O_2]_{0.5}CoO_2$. The data are well explained with the thermal activation model [$\rho(T) = \rho_0 \exp(E_a/k_B T)$ with $E_a = 7$ meV] in the higher-temperature region ($> 17$ K), see the inset.

**Conclusion**

We have successfully synthesized the parent phase of the misfit-layered cobalt-oxide family, $[M_mA_2O_{m+2}]_qCoO_2$. The new $[Sr_2O_2]_qCoO_2$ phase lacks the $M$O layer(s) so as to have a $(SrO)_2$ double layer only between adjacent $CoO_2$ layers. Even though the $(SrO)_2$ and $CoO_2$ layers in $[Sr_2O_2]_qCoO_2$ possess different symmetries (as in all other ML oxides), the phase is commensurate, *i.e.* $q = 0.5$. We have thus simplified the general $[M_mA_2O_{m+2}]_qCoO_2$ structure by reducing (*i*) the number of metal constituents and (*ii*) that of layers *per* formula unit, and moreover (*iii*) by converting the structure from incommensurate to commensurate. Nevertheless, the simplified phase yet exhibits transport-property characteristics parallel to those of the more complex derivatives of ML cobalt oxides.

**Acknowledgements**

Drs. T. Asaka (NIMS), T. Motohashi and M. Valkeapää (Tokyo Tech) are thanked for fruitful discussions. This work was supported by Grants-in-aid for Scientific Research (Nos. 15206002 and 15206071) from the Japan Society for the Promotion of Science, and also by the Nanotechnology Support Project of the MEXT, Japan.